\documentclass[twocolumn]{article} 

\usepackage{graphicx}
\usepackage{hyperref}
\usepackage{twoopt}
\usepackage{authblk}
\newcommand{\ackname}{Acknowledgements}

\begin{document}
\title{Through the eyes of a reader and science communicator:
science in the mainstream and in the genre literature of
yesterday and today}

\author[1]{Valentin D. Ivanov}
\affil[1]{\small European Southern Observatory, Karl-Schwarzschild-Str. 2,
85748 Garching bei M\"unchen, Germany; vivanov@eso.org}
\date{2023-07-11}

\maketitle

\begin{abstract}
For most writers the science is either an exotic setting 
or a source of thrilling conflict that would drive the 
story forward. For a communicator it is the other way 
around -- the science is neatly wrapped in a package of 
literary tools that make it ``invisible'' while it remains
tangible and most importantly -- it can be conveyed to 
the reader in understandable terms. There are many 
examples showing how these seemingly contradicting goals 
can complement each other successfully. I will review 
how the science was communicated by mainstream and genre 
writers of yesterday and today, and in different (not 
necessarily anglophone) cultures. I will bring forward 
the best and the worst examples that illuminate various 
astronomical concepts. Finally, I will discuss how we 
can use them both in outreach and education. Contrary to 
many similar summaries I will concentrate on some often 
overlooked mainstream literary examples, including the 
plays {\it The Physicists} by Friedrich D\"urrenmatt and
{\it Copenhagen} by Michael Frayn, the novel {\it White 
Garments} by Vl. Dudintsev and even an episode of the 
{\it Inspector Morse} TV show, featuring scientists. I 
will also mention in passing a few less well known 
genre books.
\end{abstract}

\section{Introduction}

The public image of sciences and scientists is relevant, 
especially for the fundamental science that depend on 
public support and funding. However, the sciences are 
often subject to heavy stereotyping and adversity and the 
improvement of public perception can be a daunting task.
The problem is recognized: \cite{2014PNAS.111.13593F}
polled the American public about impression of people 
according to their jobs, discriminating between ``apparent 
intent (warmth) and capability (competence).'' The outcome 
places researchers at the low-warmth/high-competence end, 
together with accountants, lawyers, engineers and chief 
executive officers. Without underestimating the importance
of these professions, we can safely assume that at least
some of them need public image improvement. Turning to 
the historic science figures, \cite{2019IJSEB...9...82C}
analyzed images of Greek scientists to trace the evolution
of how they are perceived by the public, and found
that while the public image of the Greek scientists have 
improved, it is till dominated by the traditional 
stereotypes. Studies of the public image of scientists 
and its implications for teaching of sciences in various 
cultures are also available, e.g., \cite{2003SE....12...115K}.

The depiction of sciences and scientists in arts deserves 
a special attention because of the powerful emotional 
influence that arts can exert on the public 
\cite{2007SIPIS..1... 69S,2023EI....08....14S}. Here I 
review some popular works of art that describe how
science and scientists work, and a few less well known
examples from the Eastern European add in particular from
the Bulgarian cultural lore.

\section{Scientists through the eye of the artists}

Scientists are often featured in popular arts. To remind 
a few, almost household names: Susan Calvin, Indiana Jones,
Victor Frankenstein, Dr. Strangelove, Capt. Nemo, Emmett 
Brown, Eleanor Arroway, Spock, Dr. No, Sheldon Cooper, 
Dana Scully, Dr. Kelvin, and even Sherlock Holmes. I have 
intentionally included in the list both males and females,
protagonists and antagonists, to demonstrate the diversity
of the characters, the variety of their motives and the
range of their actions. A lesser known name is that of Dr.
Kelvin, the psychologist from the novel {\it Solaris} by
Stanislaw Lem (and two motion pictures based on this book).

What brings these characters together is the analytic 
approach to the problems they face and the application of 
the scientific method; what divides them is that some use 
the scientific tools to expand our knowledge and to make 
the people's lives easier and others -- to harm people and 
society.

For writers and screenwriters invoking scientists is often 
an easy way to create all-powerful menacing figures that 
would drive the story forward without the need to invest 
much though and imagination. Further fuel into this negative 
outlook to science is provided by two notable novels 
addressing the devastating consequences that the lack of 
scientific integrity can bring -- {\it Solar} by Ian McEwan 
and {\it The White Garments} by Vladimir Dudintsev. The 
former has a modern setting and follows the life story of an 
eco-scientist who makes some very questionable decisions. 
While at its core this is a novel about integrity, the topic
is shown to exceeds the limits of the guarded and somewhat
confined scientific environment, it is shown as a problem of
world shattering proportions. The second book appeared in
the first years of the Soviet Perestroyka, when Eastern
Europe opened up and faced many issues, including to its
own past. It describes a conflict dating from the early
post-WWII years while Stalin was still in power and the
science succumbing to or rather was trying to fight a
loosing battle against ideology. The novel asked the
uncomfortable question what can one do, if the research
suggest that the world pretty much disagrees with the
established ideology and with the party line. Silence and
``internal immigration''\footnote{A term used in the former
Eastern Europe to signify a person's refusal to participate
in public life; a refusal to use social media is a modern
day analog of this behaviour.} is one answer.
Summarizing, these two books offer less than favorable
images of scientists, and they are a bit too big of a
reading to be useful topic starters that would introduce
students to ecology and biology.

Another example of a fictionalized presentation of 
scientists in a completely different environment comes from 
the Bulgarian writer Dimitar Peev (1919-1996) who published 
in 1966 a space faring novel {\it The Photon Starship}. The 
action is set in a communist Utopia and follows a sub-light 
speed trip to the nearest star system Alpha Cen. Surprisingly 
for a book of that era it contains very little ideology and
a lot of scientific speculations, some -- quite relevant 
(for how these can be used as educational topic raisers see 
\cite{2021arXiv211214702I}). The entire expedition consists 
of highly motivated scientists, so many that nobody really
stands out. This was a concession to the politics back them --
the individual did not matter, the collective and the society
ruled. However, the increasing average number of authors per
paper today may be returning us back to this idea. We can
safely say that Peev has adopted a science-optimistic stand.
His book offers insights into planet formation and stellar
binary formation, although critical comparison with the
modern understanding of these topics is needed.

In {\it Stars and Waves}
Roberto Maiolino, a professional astronomer himself, offers 
an insider's look at the life in academic environment, a 
view of the intricate research process, the motivation of 
the investigators and even some enticing description of the 
technology and methodology, both in the present day world 
leading astronomical observatory, and in big observatory 
behind the Iron Curtain from a few decades in the past. The 
novel takes the readers on a lavishly illustrated trip 
across the largest modern astronomical facilities and 
research centers. It starts like a mystery/investigation 
procedural. However, it soon enters the territory of 
international politics with hints of espionage, with detours 
towards personality studies that include big egos, hubris 
and un-moderated ambitions. Gradually some science-fictional 
elements build up, and at the end we find ourselves with a 
full fledged scifi genre work. The author does not shy away 
from some painful issues that still -- sadly -- plague the 
modern sciences, including the acceptance (or lack thereof) 
of women and ending with the outright stealing of other 
people's ideas. 

However, the novel should be taken with a big grain of salt 
as a guide to real-life university life, because the author 
has made some big concessions to build suspense and boost 
conflicts. Without revealing too much content, I can only 
tell that Laura, the graduate student protagonist, suffers 
from stage fright and have difficulties giving presentations.
Every institution that I have studied or worked at in my
career had a student seminar, where the students would 
present their work to each other, in a friendly environment, 
just to overcome this particular problem. Still, {\it Stars 
and Waves} can be a good vehicle to introduce the audience
to academic life, provided such points are addressed and
discussed. Unlike the books mentioned earlier it offers a
much more balanced view. The novel is an excellent
introduction into a discussion of a wide range of astronomy
and astrobiology topics -- from astroclimate and telescope
design to the searches for bio- and technosignatures.

The most popular TV series has been less than merciful to 
the scientists they have portrayed. {\it Inspector Morse} 
and it sequel {\it Lewis} turn Oxford into a dangerous place
with a high crime rate. The academics are frequently involved,
often shown in negative light, being involved in gambling, 
extramarital affairs and now and then -- murders. The 
protagonists often make a point of the hubris that the
scientists tend to exhibit both in their work and personal 
lives (even though the detective himself is not completely
alien to this character trait). {\it The Big Bang Theory} 
makes fun of the scientist -- without having to apply much 
imagination to the script, because the real life suffices
to make the audience laugh. At best the show evokes sympathy
to the ``poor'' scientists who are portrayed unable to coop
with the challenges of everyday life. Both these TV shows
paint negative portraits of scientists, albeit from very
different angles. {\it Inspector Morse} can hardly help as
a science intro, but {\it The Big Bang Theory} often touches
various science questions in detail.

The theatrical stage has brought us two notable examples
of science portrayals -- the plays {\it Copenhagen} and 
{\it The Physicists}. The first, written by Michael Frayn, 
is based on a real-life 
visit that Werner Heisenberg paid to Niels Bohr and his 
wife Margrethe Bohr. There are differences in the accounts 
the participants gave afterwords. Clearly, they discussed 
the atomic bomb, the implications and the ethics of 
building it. After the meeting Heisenberg went on to work 
on the Nazi nuclear program, together with Otto Hahn, Carl 
Friedrich von Weizs\"acker and other prominent physicists. 
Bohr escaped to the UK and later to the US. Overall, this
is a thoughtful piece that puts forward a picture of the
scientific word that is aware of the consequences that the
researchers will have to face in their work\footnote{The
movie {\it Oppenheimer} that premiered after the EAS 2023
addresses similar issues and it probably helped the public
to understand and sympathize with the scientists more than
any of the works mentioned here.}. While {\it Copenhagen}
does touch upon the basic nuclear physics, it is of more
help for discussions on scientific ethics.

The D\"urrenmatt play is different. It starts as a farce, 
with patients in a psychiatric yard that imagine themselves 
as Newton and Einstein, but true to the author’s love of 
surprises, they turn out to be something different yet. The
action resonates with the Cold War situation of the day and
it is not really a statement about science and scientists,
making it less relevant for our discussion.

Finally, I would like to mention four relatively recent 
collections that offer a look at the science and scientists
from insiders: {\it Diamonds in the Sky} (ed. Mike Brotherton;
2009), {\it A Kepler's Dozen: Thirteen Stories About 
Distant Worlds That Really Exist} (eds. Steve B. Howell \& 
David Lee Summers; 2013), {\it Science Fiction by Scientists} 
(ed. Mike Brotherton; 2017), and {\it Life Beyond Us} (eds. 
Julie Nov\'akov\'a, Lucas K. Law \& Susan Forest; 2023). Not 
surprisingly, the authors paint researchers in most positive 
light, as ethical problem solvers. Almost all of these were
designed to help in classroom: the stories are short and
selected to capture the attention of students; in some cases
the literary pieces are accompanied with non-fictional essays
that explain in popular terms the science that the stories
touch upon.

\section{Discussion and conclusions}

This survey how research scientist are drawn in arts is
broad, it lacks resolution and certainly does not claim
completeness
in any meaningful way. Unfortunately, even this quick
overview drives us towards a rather grim conclusion: with
a few rare exceptions the researchers are seen either as
logical but cold and hubristic beings devoid of human
qualities, who can not be relied upon, or as ill-adjusted
rueful loners. The power of fiction is to influence people,
either on conscious or on subconscious level, and here it
works toward undermining the public trust in scholarly
figures. The exceptions almost exclusively come from the
pen of a scholarly figures and reach limited audience.

Intensive outreach activities are the obvious answer to
this problem (e.g., \cite{2023arXiv231215797I}). Usually,
the outreach aims at spreading the scientific knowledge
among the general public, but others and no less important
goals are creating more science friendly environment and
putting forward a positive image of the scientists. The
outreach should bringing out the idea that doing science
is an admirable and accessible human activity, deserving
respect.

\section*{\ackname}
This is an extended write up of a contributed talk 
presented at the European Astronomical Society (EAS),  
Special Session 39 ``Sci-Art: Communicating Science 
Through Art'', held in Krak\'ow, Poland on July
10-14, 2023.

\end{document}